\documentclass[prl,twocolumn,superscriptaddress]{revtex4}

\usepackage{times}
\usepackage{amsmath}
\usepackage{amssymb}
\usepackage{graphicx}
\usepackage{pstricks}
\usepackage{wasysym}

\begin{document}

\bibliographystyle{apsrev}

\title{An extended Hubbard model with ring exchange: a route
  to a non-Abelian topological phase}



\author{Michael Freedman}
\affiliation{Microsoft Research, One Microsoft Way,
Redmond, WA 98052}

\author{Chetan Nayak}
\affiliation{Microsoft Research, One Microsoft Way,
Redmond, WA 98052}
\affiliation{Department of Physics and
  Astronomy, University of California, Los Angeles, CA 90095-1547}
\author{Kirill Shtengel}
\affiliation{Microsoft Research, One Microsoft Way,
Redmond, WA 98052}
\date{\today}

\begin{abstract}
  We propose an extended Hubbard model on a $2D$ Kagom\'{e} lattice
  with an additional ring exchange term. The particles can be either
  bosons or spinless fermions. At a special filling fraction of $1/6$
  the model is analyzed in the lowest non-vanishing order of
  perturbation theory. Such an ``undoped'' model is closely related to
  the Quantum Dimer Model. We show how to arrive at an exactly soluble
  point whose ground state is the ``$d$-isotopy'' transition point
  into a stable phase with a certain type of non-Abelian topological order.
  Near the ``special'' values, $d = 2 \cos \pi/(k+2)$, this
  topological phase has anyonic excitations closely related to
  $\text{SU}(2)$ Chern--Simons theory at level $k$.
\end{abstract}

\pacs{71.10.Hf, 71.10.Pm, 71.10.Fd}
\maketitle


Since the discovery of the fractional quantum Hall effect in 1982
\cite{Tsui82}, topological phases of electrons have been a subject of
great interest.  Many Abelian topological phases have been discovered
in the context of the quantum Hall regime \cite{DasSarma97}. More
recently, high-temperature superconductivity
\cite{Anderson87,Rokhsar-Kivelson,Kalmeyer-Laughlin,Read,
  Wen91b,Mudry94a,Fisher} and other complex materials have provided
the impetus for further theoretical studies of and experimental
searches for Abelian topological phases.  The types of microscopic
models admitting such phases are now better understood
\cite{Moessner01,Balents02,Senthil-Motrunich}.

Much less is known about non-abelian topological phases, apart from
some tantalizing hints that the quantum Hall plateau observed at
$\nu=5/2$ might correspond to such a non-abelian phase
\cite{Willett-Pan,Moore91,Greiter92,Nayak-Read-Fradkin}.  However,
non-abelian topological states, if created and controlled, would open
the door to scalable quantum computation \cite{Kitaev97,Freedman01}.
As a first step, the study of a class of topological field theories
has been reduced to combinatorial manipulations of loops on a surface
\cite{Kauffman94,Freedman03}.  A virtue of this formulation is that it
exposes a strategy for constructing microscopic physical models which
admit the corresponding phases; since Hilbert space is reduced to a
set of pictorial rules, the models should impose these rules as
energetically favorable conditions satisfied by the ground state. In
this paper, we show how this approach can be implemented.

We propose a microscopic model which has the following properties: (a)
it is an extension of the Hubbard model and, therefore, is
quasi-realistic, (b) it is soluble, (c) for certain model parameters,
it is perched at a transition point \cite{FNS04a} into a non-Abelian
topological phase relevant to quantum computation.  By
quasi-realistic, we mean that the model has short-ranged interactions
and hopping, so it is possible that the Hamiltonian of a real material
could be viewed as a small perturbation of the Hamiltonian of this
paper.  Optical lattices \cite{Greiner02}, quantum dot or Josephson
junction arrays \cite{Ioffe02} might be designed with Hamiltonians in
this general class, and these may also be promising avenues for
realizing our model.

The non-Abelian topological phases referred to in the above paragraph
are related to the doubled $\text{SU}(2)_k$ Chern-Simons theories
described in \cite{Freedman03,Freedman04a}.  These phases are
characterized by $(k+1)^2$-fold ground state degeneracy on the torus
$T^2$ and should be viewed as a natural family containing the
topological (deconfined) phase of $\text{Z}_2$ gauge theory as its
initial element, $k=1$. For $k\geq 2$ the excitations are non-Abelian.
For $k=3$ and $k\geq 5$ the excitations are computationally universal
\cite{Freedman02}.  Here, we describe the conditions which a
microscopic model should satisfy to be in such a topological phase. It
is useful to think of such a microscopic model as a lattice
regularization of a continuum model whose low energy Hilbert space may
be described as a quantum loop gas. More precisely, a state is defined
as a collection of non-intersecting loops, as discussed in
\cite{Freedman04a,FNS03b,FNS04a}.  A Hamiltonian acting on such state
can do the following: (i) the loops can be continuously deformed -- we
will call this ``move'' an isotopy move; (ii) a small loop can be
created or annihilated -- the combined effect of this move and the
isotopy move has been dubbed
`$d$-isotopy'\cite{Freedman03,Freedman04a,FNS04a}; (iii) finally, when
exactly $k+1$ strands come together in some local neighborhood, the
Hamiltonian can cut them and reconnect the resulting ``loose ends''
pairwise so that the newly-formed loops are still non-intersecting.
More specifically, in order for this model to be in a topological
phase, the ground state of this Hamiltonian should be a superposition
of all such pictures with the additional requirements that (i) if two
pictures can be continuously deformed into each other, they enter the
ground state superposition with the same weight; (ii) the amplitude of
a picture with an additional loop is $d$ times that of a picture
without such loop; (iii) this superposition is annihilated by the
application of the Jones--Wenzl (JW) projector that acts locally by
reconnecting $k+1$ strands. Readers interested in the details are
referred to \cite{Freedman03,Freedman04a,FNS03b} and references
therein. In this paper, we focus on the first two conditions, which
place the system at a transition point into the desired phase(s)
\cite{FNS04a}; our purpose is to construct a Hamiltonian which
enforces $d$-isotopy for its ground state(s).

\begin{figure}[hbt]
\includegraphics[width=3in]{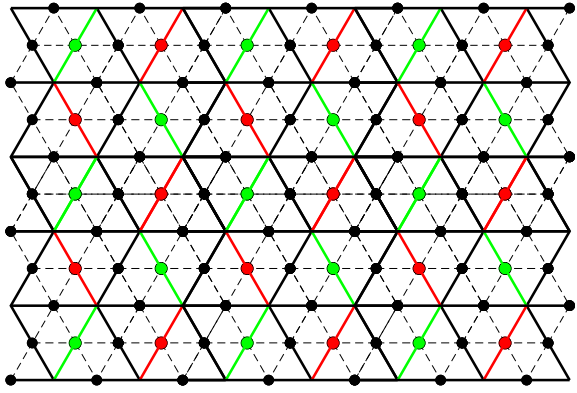}
\caption{Solid dots and dashed lines represent sites and bonds of
  the Kagom\'{e} lattice $\mathcal{K}$
  with the special sublattices $\mathcal{R}$ (red) and $\mathcal{G}$
  (green). Solid lines define the surrounding triangular lattice.}
\label{fig:Kagome}
\end{figure}

Our proposed model is defined on the Kagom\'{e} lattice
shown in Fig.~\ref{fig:Kagome}. The sites of the lattice are not
completely equivalent, in particular we choose two special sublattices
- $\mathcal{R}$ (red) and $\mathcal{G}$ (green) whose significance
will be discussed later. The Hamiltonian is given by:
\begin{multline}
  \label{eq:Hubbard}
  H =  \sum_i \mu_i n_i +  U_0 \sum_i n^2_i
  +  U\!\!\!\sum_{(i,j)\in \hexagon}\!\! n_i n_j
  + \!\!\!\!\sum_{(i,j)\in \bowtie, \notin \hexagon }\!\!\!
  V_{i j} n_i n_j\\
   - \sum_{\langle i,j \rangle} t_{ij}
  (c^\dag_i c_j + c^\dag_j c_i ) + T_\text{R}.
\end{multline}
Here $n_i \equiv c_i^\dag c_i$ is the occupation number on site $i$,
$\mu_i$ is the corresponding chemical potential. $U_0$ is the usual
onsite Hubbard energy $U_0$ (clearly superfluous for spinless
fermions). $U$ is a (positive) Coulomb penalty for having two
particles on the same hexagon while $V_{ij}$ represent a penalty for
two particles occupying the opposite corners of ``bow ties'' (in other
words, being next-nearest neighbors on one of the straight lines). We
allow for the possibility of inhomogeneity so not all $V_{ij}$ are
assumed equal.  Specifically, define $v^{c}_{ab}= V_{ij}$ where $a$ is
the color of site $(i)$, $b$ is the color of $(j)$, and $c$ is the
color of the site between them. In the lattice in
Fig.~\ref{fig:Kagome} we have, possibly distinct, $v^{g}_{bb}$,
$v^{b}_{bb}$, $v^{g}_{bg}$, $v^{b}_{rb}$, and $v^{b}_{rg}$, where $r
\in \mathcal{R}$, $g \in \mathcal{G}$, and $b \in \mathcal{B}=
\mathcal{K} \diagdown (\mathcal{R} \cup \mathcal{G})$.  $t_{ij}$ is
the usual nearest-neighbor tunnelling amplitude which is also assumed
to depend only on the color of the environment: $t_{ij}\equiv
t^{c}_{ab}$ where $c$ now refers to the color of the third site in a
triangle.  Finally, we include $T_\text{R}$ -- a four-particle ring
exchange term whose exact form will be specified later.  $T_\text{R}$
is added to the Hamiltonian on an \emph{ad hoc} basis to allow
correlated multi-particle hops.  Ring exchange terms can be justified
semiclassically \cite{Roger84} and they do indeed appear in in such
physical systems as spin systems \cite{Herring62,MacDonald88}, solid
${}^{3}$He \cite{Thouless65} and Wigner crystals
\cite{Chakravarty}. Of course, small ring terms can arise
perturbatively along the lines of \cite{MacDonald88}, e.g. a
four-particle move occurs at order 4.

The onsite Hubbard energy $U_0$ is considered to be the biggest energy
in the problem, and we shall set it to infinity, thereby restricting
our attention to the low-energy manifold with sites either unoccupied
or singly-occupied. The rest of the energies satisfy the following
relations: $U\gg t_{ij},V_{ij}, \mu_{i}$; we shall be more specific
about relations between various $t_{ij}$'s, $V_{ij}$'s and $\mu_i$'s
later.

The ``undoped'' system corresponds to the filling fraction 1/6 (i.e.
$N_p\equiv\sum_i n_i = N/6$, where $N$ is the number of sites in the
lattice).  The lowest-energy band then consists of configurations in
which there is exactly one particle per hexagon, hence all $U$-terms
are set to zero.  These states are easier to visualize if we consider
a triangular lattice $\mathcal{T}$ whose sites coincide with the
centers of hexagons of $\mathcal{K}$.  ($\mathcal{K}$ is a
\emph{surrounding} lattice for $\mathcal{T}$.)  Then a particle on
$\mathcal{K}$ is represented by a dimer on $\mathcal{T}$ connecting
the centers of two adjacent hexagons of $\mathcal{K}$.  The condition
of one particle per hexagon translates into the requirement that no
dimers share a site.  In the 1/6-filled case this low-energy manifold
coincides with the set of all dimer coverings (perfect matchings) of
$\mathcal{T}$.  The ``red'' bonds of $\mathcal{T}$ (the ones
corresponding to the sites of sublattice $\mathcal{R}$) themselves
form one such dimer covering, a so-called ``staggered configuration''.
This particular covering is special: it contains no ``flippable
plaquettes'', or rhombi with two opposing sides occupied by dimers
(see Fig.~\ref{fig:Kagome}).

So henceforth particles live on \emph{bonds} of the triangular lattice
(Fig.~\ref{fig:Kagome}) and are represented as dimers \footnote{It is
  important that the triangular lattice is that it is \emph{not}
  bipartite. On the edges of a bipartite lattice, our models will have
  an additional, undesired, conserved quantity (integral winding
  numbers, which are inconsistent with the JW projectors for $k>2$),
  so the triangular lattice gives the simplest realization.}. In
particular, a particle hop corresponds to a dimer ``pivoting'' by
$60^\circ$ around one of its endpoints, $V_{ij}=v^{c}_{ab}$ is now a
potential energy of two parallel dimers on two opposite sides of a
rhombus (with $c$ being the color of its short diagonal).  It is clear
that our model is in the same family as the quantum dimer model
\cite{Rokhsar-Kivelson}, which has recently been shown to have an
Abelian topological phase on the triangular lattice \cite{Moessner01}
which, corresponds to $k=1$, or $d=1$.  Here, we show how other values
of $k$ can be obtained.

The goal now is to derive the effective Hamiltonian acting on this
low-energy manifold represented by all possible dimer coverings of
$\mathcal{T}$.  Our analysis is perturbative in ${t}/{U}=: \epsilon$.
The initial, unperturbed ground state manifold for $U_0=\infty$, $U$
large and positive, all $t_{ij},\; V_{ij}=0$ and all $\mu_i$ equal is
spanned by the dimerizations $\mathcal{D}$ of the triangular lattice
$\mathcal{T}$.
As we gradually turn on the $t$'s, $v$'s, and $T_{\text{R}}$, we shall
see what equations they should satisfy so that the effective
Hamiltonian on $\mathcal{D}$ has the desired $d$-isotopy space as its
ground state(s).

Since a single tunnelling event in $\mathcal{D}$ always leads to dimer
``collisions'' (two dimers sharing an endpoint) with energy penalty
$U$, the lowest order at which the tunnelling processes contribute to
the effective low-energy Hamiltonian is 2.  At this order, the
tunnelling term leads to two-dimer ``plaquette flips'' as well as
renormalization of bare onsite potentials $\mu_i$'s due to dimers
pivoting out of their positions and back.  We always recompute bare
potentials $\mu_i$'s to maintain equality up to errors $\mathcal{O}
(\epsilon^3 )$ among the renormalized $\widetilde{\mu}_i$'s.  This
freedom to engineer the chemical potential landscape to balance
kinetic energy is essential to finding an exactly soluble point.

Let us pause and discuss the connection between our quantum
dimer model and a desired topological phase . It is an old idea (see
e.g. \cite{Sutherland88}) to turn a dimerization (perfect matching)
$\mathcal{J}$ into a collection of loops by using a background
dimerization $\mathcal{R}$ to form a `transition graph' $\mathcal{R}
\cup \mathcal{J}$.  It turns out that fixing $\mathcal{R}$ as in
Fig.~\ref{fig:Kagome}, without small rhombi with two opposite sides
red, as the preferred background dimerization we obtain the fewest
equations in and also achieve ergodicity \cite{Kenyon96} under a small
set of moves.  Unlike in the usual case, the background dimerization
$\mathcal{R}$ is not merely a guide for the eyes, it is
\emph{physically} distinguished: the chemical potentials and
tunnelling amplitudes are different for bonds of different color.

\begin{figure}[htb]
\includegraphics[width=2.75in]{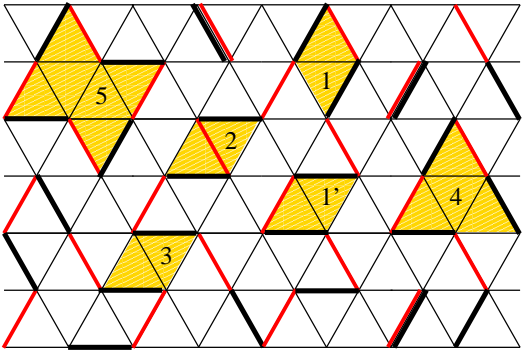}
\caption{
  Overlap of a dimer covering of $\mathcal{T}$ (shown in thick black)
  with the red covering corresponding to the special sublattice
  $\mathcal{R}$.  Shaded plaquettes correspond to various dimer moves
  described in the text. Green sublattice is not shown.}
\label{fig:Overlap}
\end{figure}

Let us list here the elementary dimer moves that preserve the proper
dimer covering condition:\\
(i) {Plaquette (rhombus) flip -- this is a two-dimer move around a
    rhombus made of two lattice triangles. Depending on whether a
    ``red'' bond forms a side of such a rhombus, its diagonal, or is
    not found there at all, the plaquettes are referred to,
    respectively, as type 1 (or 1'), 2, or 3 (see
    Fig.~\ref{fig:Overlap}). The distinction between plaquettes of
    type 1 and 1' is purely directional: diagonal bonds in plaquettes
    of type 1 are horizontal, for type 1' they are not. This
    distinction is necessary since our Hamiltonian breaks the
    rotational symmetry of a triangular (or Kagom\'{e}) lattice.}\\
(ii) {Triangle move -- this is a three-dimer move around a triangle
    made of four elementary triangles. One such ``flippable'' triangle
    is labelled 4 in Fig.~\ref{fig:Overlap}.}\\
(iii) {Bow tie move -- this is a four-dimer move around a ``bow tie''
    made of six elementary triangles. One such ``flippable'' bow tie
    is labelled 5 in Fig.~\ref{fig:Overlap}.}

To make each of the above moves possible, the actual dimers and
unoccupied bonds should alternate around a corresponding shape.
Notice that for both triangle and bow tie moves we chose to depict the
cases when the maximal possible number of ``red'' bonds participate in
their making (2 and 4 respectively).  Note that there are no
alternating red/black rings of fewer than 8 lattice bonds (occupied by
at most 4 non-colliding dimers). Ring moves only occur when red and
black dimers alternate; the triangle labelled 4 in
Fig.~\ref{fig:Overlap} does not have a ring exchange term associated
with it, but the bow tie labelled 5 does:
  \begin{eqnarray}
    \label{eq:ring-exchange}
    T_{\text{R}}= a\;  \Bigg|
    \lambda \;
    \pspicture[-0.05](.8,0.69282)
    \psset{linewidth=2pt}
    \psline[linecolor=black](0,0.34641)(0.4,0.34641)
    \psline[linecolor=red](0.4,0.34641)(0.8,0.34641)
    \psline[linecolor=black](0.8,0.34641)(0.6,0)
    \psline[linecolor=red](0.6,0)(0.8,-0.34641)
    \psline[linecolor=black](0.8,-0.34641)(0.4,-0.34641)
    \psline[linecolor=red](0.4,-0.34641)(0,-0.34641)
    \psline[linecolor=black](0,-0.34641)(0.2,0)
    \psline[linecolor=red](0.2,0)(0,0.34641)
    \psline[linecolor=black,linestyle=dotted,linewidth=1pt](0.2,0)(0.6,0)
    \pspolygon[linecolor=black,linestyle=dotted,linewidth=1pt](0.2,0)
    (0.4,0.34641)(0.6,0)(0.4,-0.34641)
    \endpspicture
    -
    \pspicture[-0.05](.8,0.69282)
    \psset{linewidth=1.5pt}
    \psline[linecolor=black,linestyle=dotted,linewidth=1pt](0,0.34641)
    (0.4,0.34641)
    \psline[linecolor=red](0.4,0.34641)(0.8,0.34641)
    \psline[linecolor=black](0.4,0.38641)(0.8,0.38641)
    \psline[linecolor=black,linestyle=dotted,linewidth=1pt](0.8,0.34641)
    (0.6,0)
    \psline[linecolor=red](0.6,0)(0.8,-0.34641)
    \psline[linecolor=black](0.63,0.03)(0.83,-0.31641)
    \psline[linecolor=black,linestyle=dotted,linewidth=1pt](0.8,-0.34641)
    (0.4,-0.34641)
    \psline[linecolor=red](0.4,-0.34641)(0,-0.34641)
    \psline[linecolor=black](0.4,-0.30641)(0,-0.30641)
    \psline[linecolor=black,linestyle=dotted,linewidth=1pt](0,-0.34641)
    (0.2,0)
    \psline[linecolor=red](0.2,0)(0,0.34641)
    \psline[linecolor=black](0.23,0.03)(0.03,0.37641)
    \psline[linecolor=black,linestyle=dotted,linewidth=1pt](0.2,0)(0.6,0)
    \pspolygon[linecolor=black,linestyle=dotted,linewidth=1pt](0.2,0)
    (0.4,0.34641)(0.6,0)(0.4,-0.34641)
    \endpspicture
    \Bigg\rangle
    \Bigg\langle
    \lambda \;
    \pspicture[-0.05](.8,0.69282)
    \psset{linewidth=2pt}
    \psline[linecolor=black](0,0.34641)(0.4,0.34641)
    \psline[linecolor=red](0.4,0.34641)(0.8,0.34641)
    \psline[linecolor=black](0.8,0.34641)(0.6,0)
    \psline[linecolor=red](0.6,0)(0.8,-0.34641)
    \psline[linecolor=black](0.8,-0.34641)(0.4,-0.34641)
    \psline[linecolor=red](0.4,-0.34641)(0,-0.34641)
    \psline[linecolor=black](0,-0.34641)(0.2,0)
    \psline[linecolor=red](0.2,0)(0,0.34641)
    \psline[linecolor=black,linestyle=dotted,linewidth=1pt](0.2,0)(0.6,0)
    \pspolygon[linecolor=black,linestyle=dotted,linewidth=1pt](0.2,0)
    (0.4,0.34641)(0.6,0)(0.4,-0.34641)
    \endpspicture
    -
    \pspicture[-0.05](.8,0.69282)
    \psset{linewidth=1.5pt}
    \psline[linecolor=black,linestyle=dotted,linewidth=1pt](0,0.34641)
    (0.4,0.34641)
    \psline[linecolor=red](0.4,0.34641)(0.8,0.34641)
    \psline[linecolor=black](0.4,0.38641)(0.8,0.38641)
    \psline[linecolor=black,linestyle=dotted,linewidth=1pt](0.8,0.34641)
    (0.6,0)
    \psline[linecolor=red](0.6,0)(0.8,-0.34641)
    \psline[linecolor=black](0.63,0.03)(0.83,-0.31641)
    \psline[linecolor=black,linestyle=dotted,linewidth=1pt](0.8,-0.34641)
    (0.4,-0.34641)
    \psline[linecolor=red](0.4,-0.34641)(0,-0.34641)
    \psline[linecolor=black](0.4,-0.30641)(0,-0.30641)
    \psline[linecolor=black,linestyle=dotted,linewidth=1pt](0,-0.34641)
    (0.2,0)
    \psline[linecolor=red](0.2,0)(0,0.34641)
    \psline[linecolor=black](0.23,0.03)(0.03,0.37641)
    \psline[linecolor=black,linestyle=dotted,linewidth=1pt](0.2,0)(0.6,0)
    \pspolygon[linecolor=black,linestyle=dotted,linewidth=1pt](0.2,0)
    (0.4,0.34641)(0.6,0)(0.4,-0.34641)
    \endpspicture
    \; \Bigg|.
  \end{eqnarray}

Here is the correspondence between the previous smooth discussion and
rhombus flips relating dimerizations $\mathcal{J}$ of $\mathcal{T}$.
Our surface is now a planar domain with, possibly, periodic boundary
conditions (a torus). A collection of loops is generated by
$\mathcal{R} \cup \mathcal{J}$ (with the convention that the dimers of
$\mathcal{R} \cap \mathcal{J}$ be consider as length 2 loops or
bigons). What about isotopy?  Move 2 certainly is an isotopy from
$\mathcal{R} \cup \mathcal{J}$ to $\mathcal{R} \cup \mathcal{J}'$ but
 by itself, it does almost nothing.  It is impossible to build
up large moves from type 2 alone.  So it is a peculiarity of the rhombus
flips that we have no good analog of isotopy alone but instead go
directly to $d-$isotopy.
We should impose the
following relations associated with
moves of type 1 (1'):
\begin{eqnarray}
  d\; \Psi \left(
    \pspicture[0.0](.6,0.34641)
    \psset{linewidth=2pt}
    \psline[linecolor=black](0,-0.1)(0.2,0.24641)
    \psline[linecolor=red](0.2,0.24641)(0.6,0.24641)
    \psline[linecolor=black](0.6,0.24641)(0.4,-0.1)
    \psline[linecolor=black,linestyle=dotted,linewidth=1pt](0,-0.1)(0.4,-0.1)
    \psline[linecolor=black,linestyle=dotted,linewidth=1pt](0.2,0.24641)
    (0.4,-0.1)
    \endpspicture
  \right)
  - \Psi \left(
    \pspicture[0.0](.6,0.34641)
    \psset{linewidth=2pt}
    \psline[linecolor=black,linestyle=dotted,linewidth=1pt](0,-0.1)
    (0.2,0.24641)
    \psline[linecolor=red,linewidth=1.5pt](0.2,0.24641)(0.6,0.24641)
    \psline[linecolor=black,linewidth=1.5pt](0.2,0.28641)(0.6,0.28641)
    \psline[linecolor=black,linestyle=dotted,linewidth=1pt](0.6,0.24641)
    (0.4,-0.1)
    \psline[linecolor=black](0,-0.1)(0.4,-0.1)
    \psline[linecolor=black,linestyle=dotted,linewidth=1pt](0.2,0.24641)
    (0.4,-0.1)
    \endpspicture
  \right)
  =0
  \label{dimer_rel_1b}
\end{eqnarray}
since we pass from zero to one loop in (\ref{dimer_rel_1b}).
Additionally, the ring exchange term (\ref{eq:ring-exchange})
annihilates the superposition of one and four loops; we therefore
require that $\lambda = d^{3}$.

Having stated our goal, we now derive the effective Hamiltonian
$\tilde{H}: \mathcal{D}\to \mathcal{D}$ on the span of dimerizations.
The derivation is perturbative to the second order in $\epsilon$ where
$\epsilon=t^{r}_{bb}/U=t^{b}_{gb}/U$. Additionally, $t^{b}_{rb}/U =
c_0 \epsilon$ where $c_0$ is a positive constant, while
$t^{g}_{bb}={o}(\epsilon)$ and can be neglected in the second-order
calculations. (In the absence of a magnetic field all $t$'s can be
made real and hence symmetric with respect to their lower indices.
Also, we set $U =1$ for notational convenience.)  We account for all
second-order processes, i.e. those processes that take us out of
$\mathcal{D}$ and then back to $\mathcal{D}$. These amount to
off-diagonal (hopping) processes -- ``plaquette'' flips'' or ``rhombus
moves'' -- as well as diagonal ones (potential energy) in which a
dimer pivots out and then back into its original position.  The latter
processes lead to renormalization of the bare onsite potentials
$\mu_i$, which we have adjusted so that all renormalized potentials
$\tilde\mu_i$ are equal up to corrections $\mathcal{O}(\epsilon^3)$.
The non-constant part of the effective Hamiltonian comes from the
former processes and can be written in the form: $\tilde{H}=
\sum_{\mathcal{I},\mathcal{J}} \left( \tilde{H}_{\mathcal{I}
    \mathcal{J}} \otimes \mathbb{I}\right) \Delta_{\mathcal{I}
  \mathcal{J}}$ where $\tilde{H}_{\mathcal{I}\mathcal{J}}$ is a
$2\times 2$ matrix corresponding to a dimer move in the
two-dimensional basis of dimer configurations connected by this move.
$\Delta_{\mathcal{I} \mathcal{J}}=1$ if the dimerizations
${\mathcal{I},\mathcal{J}}\in \mathcal{D}$ are connected by an allowed
move, $\Delta_{\mathcal{I} \mathcal{J}}=0$ otherwise.  Therefore it
suffices to specify these $2\times 2$ matrices $\tilde{H}_{\mathcal{I}
  \mathcal{J}}$ for the off-diagonal processes.  For moves of types
(1)--(3), they are given below:
\begin{subequations}
  \label{eq:regular_moves}
  \begin{eqnarray}
    \label{eq:move_1}
    \tilde{H}^{(1)}=
    \begin{pmatrix}
      v^{b}_{gb}&  -2 t^{b}_{rb} t^{b}_{gb}    \\
      -2 t^{b}_{rb} t^{b}_{gb}&  v^{b}_{rb}
    \end{pmatrix}
    =
    \begin{pmatrix}
      v^{b}_{gb}&  -2 c_0 \epsilon^{2}    \\
      -2 c_0 \epsilon^{2}&  v^{b}_{rb}
    \end{pmatrix},
  \end{eqnarray}
  \begin{eqnarray}
    \label{eq:move_1_prime}
    \tilde{H}^{(1')}=
    \begin{pmatrix}
      v^{b}_{bb}&  -2 t^{b}_{rb} t^{b}_{gb}    \\
      -2 t^{b}_{rb} t^{b}_{gb}&  v^{b}_{rg}
  \end{pmatrix}
  =
  \begin{pmatrix}
    v^{b}_{bb}&  -2 c_0 \epsilon^2    \\
    -2 c_0 \epsilon^2 &  v^{b}_{rb}
  \end{pmatrix},
\end{eqnarray}
\begin{eqnarray}
  \label{eq:move_2}
  \tilde{H}^{(2)}=
  \begin{pmatrix}
    v^{r}_{bb}&  -2 (t^{r}_{bb})^{2}    \\
    -2 (t^{r}_{bb})^{2} &  v^{r}_{bb}
  \end{pmatrix}
  =
  \begin{pmatrix}
    v^{r}_{bb}&  -2 \epsilon^{2}   \\
    -2 \epsilon^{2}&  v^{r}_{bb}
  \end{pmatrix},
\end{eqnarray}
\begin{eqnarray}
  \label{eq:move_3}
  \tilde{H}^{(3)}= \begin{pmatrix}
    v^{g}_{bb}&  -2 (t^{g}_{bb})^{2}    \\
    -2 (t^{g}_{bb})^{2} &  v^{g}_{bb}
  \end{pmatrix}
  =
  \begin{pmatrix}
    v^{g}_{bb} &  0   \\
    0 & v^{g}_{bb}
  \end{pmatrix}.
\end{eqnarray}
\end{subequations}
We can now tune $\tilde{H}$ to the ``small
loop'' value $d$.
We require
$\tilde{H}^{(1)}= \tilde{H}^{(1')} \propto \left(
\begin{smallmatrix}
  d &  -1  \\
  -1 &  d^{-1}
\end{smallmatrix}
\right) $ as these moves change the number of small loops by one
(cf. Eq.~(\ref{dimer_rel_1b})).
Since a move of type 2 is just an isotopy move, we require
$\tilde{H}^{(2)} \propto \left(
\begin{smallmatrix}
  1&  -1   \\
  -1& 1
\end{smallmatrix}\right)
$.  Finally, $\tilde{H}^{(3)}= 0$ provided $k>1$, since it represents
a ``surgery'' on two strands not allowed for $k>1$.  (For $k=1$, on
the other hand, $\tilde{H}^{(3)}\propto \left(
  \begin{smallmatrix}
    1&  -1   \\
    -1& 1
  \end{smallmatrix}\right)
$.)  At level $k=1$ configurations which differ by such a surgery
should have equal coefficients in any ground state vector $\Psi$ while
at levels $k>1$ no such relation should be imposed.  Thus, for $k>1$
the matrix relations (\ref{eq:move_1}-\ref{eq:move_3}) yield equations
in the model parameters:
\begin{subequations}
  \label{eq:param_cond_A}
  \begin{eqnarray}
    \text{Types } (1) \& (1'):  &  v^b_{gb} = v^b_{bb} = 2dc_0 \epsilon^2
    \\
    { } & \text{and}\quad v^b_{rb} = v^b_{rg}= 2d^{-1} c_0 \epsilon^2
    \\
    \text{Types } (2) \& (3):  & v^r_{bb} = 2\epsilon^2
    \quad \text{and} \quad
    v^g_{bb} = 0
  \end{eqnarray}
\end{subequations}

We have already assumed that the Hamiltonian has a bare ring exchange
term, $T_{\text{R}}$ given by Eq.~(\ref{eq:ring-exchange}) or, in
matrix form, $T_{\text{R}}= a \left(
  \begin{smallmatrix}
     \lambda^2 & -\lambda  \\
    -\lambda  &  1
  \end{smallmatrix} \right)$ where $\lambda=d^3$ according to the discussion
after Eq.~(\ref{dimer_rel_1b}).  Additionally, we would want the
off-diagonal elements of $T_{\text{R}}$, $- a \lambda$ to be of order
$\epsilon^2$ thus making sure that this ring exchange dominates all
other ring exchanges that will appear in the higher orders of
perturbation theory.  Along with Eqs.~(\ref{eq:param_cond_A}), these
conditions place our model at the soluble point characterized by
$d$-isotopy.  We remark that additional freedom in defining
$T_{\text{R}}$ can be gained by exploiting the the ambiguity of
whether a bigon should be considered a loop or not, as discussed in
\cite{FNS03b}. In particular, this allows one to make the diagonal
elements of $T_{\text{R}}$ equal.

This construction shows how an extended Hubbard model with an
additional ring exchange term (or the equivalent Quantum Dimer Model)
can be tuned to the $d$-isotopy state(s). As discussed earlier, they
satisfy two of the three conditions which define a class of stable,
gapped topological phases which are centered about the special values
$d=2\cos(\pi/k+2)$.  The next step is to understand how perturbations
can push the system (by implementing the JW projectors) into these
phases. Our simplest candidate for a ``universal quantum computer'' is
associated with $d=({1+\sqrt{5}})/{2}$.

\begin{acknowledgments}
  The authors are grateful to D.~Jetchev for kindly providing a
  computer program for exploring dimer dynamics. It is a pleasure to
  acknowledge helpful discussions with M.P.A. Fisher, S.~Kivelson,
  S.~Sondhi, K.~Walker, and Z.~Wang and the hospitality of the Aspen
  Center for Physics where a part of this paper was completed. C.N.
  acknowledges the support of the NSF under grant DMR-9983544 and the
  Alfred P.  Sloan Foundation.
\end{acknowledgments}


\begin{thebibliography}{38}
\expandafter\ifx\csname natexlab\endcsname\relax\def\natexlab#1{#1}\fi
\expandafter\ifx\csname bibnamefont\endcsname\relax
  \def\bibnamefont#1{#1}\fi
\expandafter\ifx\csname bibfnamefont\endcsname\relax
  \def\bibfnamefont#1{#1}\fi
\expandafter\ifx\csname citenamefont\endcsname\relax
  \def\citenamefont#1{#1}\fi
\expandafter\ifx\csname url\endcsname\relax
  \def\url#1{\texttt{#1}}\fi
\expandafter\ifx\csname urlprefix\endcsname\relax\def\urlprefix{URL }\fi
\providecommand{\bibinfo}[2]{#2}
\providecommand{\eprint}[2][]{\url{#2}}

\bibitem[{\citenamefont{Tsui et~al.}(1982)\citenamefont{Tsui, Stormer, and
  Gossard}}]{Tsui82}
\bibinfo{author}{\bibfnamefont{D.~C.} \bibnamefont{Tsui}},
  \bibinfo{author}{\bibfnamefont{H.~L.} \bibnamefont{Stormer}},
  \bibnamefont{and} \bibinfo{author}{\bibfnamefont{A.~C.}
  \bibnamefont{Gossard}}, \bibinfo{journal}{Phys. Rev. Lett.}
  \textbf{\bibinfo{volume}{48}}, \bibinfo{pages}{1559} (\bibinfo{year}{1982}).

\bibitem[{\citenamefont{Das~Sarma and Pinczek}(1997)}]{DasSarma97}
  \bibinfo{author}{\bibfnamefont{S.}~\bibnamefont{Das~Sarma}}
  \bibnamefont{and}
  \bibinfo{author}{\bibfnamefont{A.}~\bibnamefont{Pinczek}},
  \emph{\bibinfo{title}{Perspectives in quantum Hall effects : novel
      quantum liquids in low-dimensional semiconductor structures}}
  (\bibinfo{publisher}{Wiley}, \bibinfo{address}{New York},
  \bibinfo{year}{1997}).

\bibitem[{\citenamefont{Anderson}(1987)}]{Anderson87}
\bibinfo{author}{\bibfnamefont{P.~W.} \bibnamefont{Anderson}},
  \bibinfo{journal}{Science} \textbf{\bibinfo{volume}{235}},
  \bibinfo{pages}{1196} (\bibinfo{year}{1987}).

\bibitem[{\citenamefont{Kivelson et~al.}(1987)\citenamefont{Kivelson,
      Rokhsar, and Sethna}}]{Rokhsar-Kivelson}
  \bibinfo{author}{\bibfnamefont{S.~A.} \bibnamefont{Kivelson}},
  \bibinfo{author}{\bibfnamefont{D.~S.} \bibnamefont{Rokhsar}},
  \bibnamefont{and} \bibinfo{author}{\bibfnamefont{J.~P.}
    \bibnamefont{Sethna}}, \bibinfo{journal}{Phys. Rev. B}
  \textbf{\bibinfo{volume}{35}}, \bibinfo{pages}{8865}
  (\bibinfo{year}{1987});
  \bibinfo{author}{\bibfnamefont{D.~S.} \bibnamefont{Rokhsar}}
  \bibnamefont{and} \bibinfo{author}{\bibfnamefont{S.~A.}
    \bibnamefont{Kivelson}}, \bibinfo{journal}{Phys. Rev. Lett.}
  \textbf{\bibinfo{volume}{61}}, \bibinfo{pages}{2376}
  (\bibinfo{year}{1988}).

\bibitem[{\citenamefont{Kalmeyer and
      Laughlin}(1987)}]{Kalmeyer-Laughlin}
  \bibinfo{author}{\bibfnamefont{V.}~\bibnamefont{Kalmeyer}}
  \bibnamefont{and} \bibinfo{author}{\bibfnamefont{R.~B.}
    \bibnamefont{Laughlin}}, \bibinfo{journal}{Phys. Rev. Lett.}
  \textbf{\bibinfo{volume}{59}}, \bibinfo{pages}{2095}
  (\bibinfo{year}{1987});
  \bibinfo{author}{\bibfnamefont{R.~B.} \bibnamefont{Laughlin}},
  \bibinfo{journal}{Phys. Rev. Lett.} \textbf{\bibinfo{volume}{60}},
  \bibinfo{pages}{2677} (\bibinfo{year}{1988}{\natexlab{a}});
\bibinfo{author}{\bibfnamefont{R.~B.} \bibnamefont{Laughlin}},
  \bibinfo{journal}{Science} \textbf{\bibinfo{volume}{242}},
  \bibinfo{pages}{525} (\bibinfo{year}{1988}{\natexlab{b}});
  \bibinfo{author}{\bibfnamefont{A.}~\bibnamefont{Fetter}},
  \bibinfo{author}{\bibfnamefont{C.}~\bibnamefont{Hanna}},
  \bibnamefont{and}
  \bibinfo{author}{\bibfnamefont{R.}~\bibnamefont{Laughlin}},
  \bibinfo{journal}{Phys. Rev. B} \textbf{\bibinfo{volume}{39}},
  \bibinfo{pages}{9679} (\bibinfo{year}{1989});
\bibinfo{author}{\bibfnamefont{Y.}~\bibnamefont{Chen}},
  \bibinfo{author}{\bibfnamefont{F.}~\bibnamefont{Wilczek}},
  \bibinfo{author}{\bibfnamefont{E.}~\bibnamefont{Witten}}, \bibnamefont{and}
  \bibinfo{author}{\bibfnamefont{B.}~\bibnamefont{Halperin}},
  \bibinfo{journal}{Int. J. Mod. Phys. B} p. \bibinfo{pages}{1001}
  (\bibinfo{year}{1989}).

\bibitem[{\citenamefont{Read and Chakraborty}(1989)}]{Read}
\bibinfo{author}{\bibfnamefont{N.}~\bibnamefont{Read}} \bibnamefont{and}
  \bibinfo{author}{\bibfnamefont{B.}~\bibnamefont{Chakraborty}},
  \bibinfo{journal}{Phys. Rev. B} \textbf{\bibinfo{volume}{40}},
  \bibinfo{pages}{7133} (\bibinfo{year}{1989});
\bibinfo{author}{\bibfnamefont{N.}~\bibnamefont{Read}} \bibnamefont{and}
  \bibinfo{author}{\bibfnamefont{S.}~\bibnamefont{Sachdev}},
  \bibinfo{journal}{Phys. Rev. Lett.} \textbf{\bibinfo{volume}{66}},
  \bibinfo{pages}{1773} (\bibinfo{year}{1991}{\natexlab{a}});
\bibinfo{author}{\bibfnamefont{N.}~\bibnamefont{Read}} \bibnamefont{and}
  \bibinfo{author}{\bibfnamefont{S.}~\bibnamefont{Sachdev}},
  \bibinfo{journal}{Int. J. Mod. Phys. B} \textbf{\bibinfo{volume}{5}},
  \bibinfo{pages}{219} (\bibinfo{year}{1991}{\natexlab{b}}).

\bibitem[{\citenamefont{Wen}(1991)}]{Wen91b}
\bibinfo{author}{\bibfnamefont{X.~G.} \bibnamefont{Wen}},
  \bibinfo{journal}{Phys. Rev. B} \textbf{\bibinfo{volume}{44}},
  \bibinfo{pages}{2664} (\bibinfo{year}{1991}).

\bibitem[{\citenamefont{Mudry and Fradkin}(1994)}]{Mudry94a}
\bibinfo{author}{\bibfnamefont{C.}~\bibnamefont{Mudry}} \bibnamefont{and}
  \bibinfo{author}{\bibfnamefont{E.}~\bibnamefont{Fradkin}},
  \bibinfo{journal}{Phys. Rev. B} \textbf{\bibinfo{volume}{49}},
  \bibinfo{pages}{5200} (\bibinfo{year}{1994}).

\bibitem[{\citenamefont{Balents et~al.}(1998)\citenamefont{Balents, Fisher, and
  Nayak}}]{Fisher}
\bibinfo{author}{\bibfnamefont{L.}~\bibnamefont{Balents}},
  \bibinfo{author}{\bibfnamefont{M.~P.~A.} \bibnamefont{Fisher}},
  \bibnamefont{and} \bibinfo{author}{\bibfnamefont{C.}~\bibnamefont{Nayak}},
  \bibinfo{journal}{Int. J. Mod. Phys. B} \textbf{\bibinfo{volume}{12}},
  \bibinfo{pages}{1033} (\bibinfo{year}{1998});
\bibinfo{author}{\bibfnamefont{T.}~\bibnamefont{Senthil}} \bibnamefont{and}
  \bibinfo{author}{\bibfnamefont{M.~P.~A.} \bibnamefont{Fisher}},
  \bibinfo{journal}{Phys. Rev. B} \textbf{\bibinfo{volume}{62}},
  \bibinfo{pages}{7850} (\bibinfo{year}{2000}).

\bibitem[{\citenamefont{Moessner and Sondhi}(2001)}]{Moessner01}
\bibinfo{author}{\bibfnamefont{R.}~\bibnamefont{Moessner}} \bibnamefont{and}
  \bibinfo{author}{\bibfnamefont{S.~L.} \bibnamefont{Sondhi}},
  \bibinfo{journal}{Phys. Rev. Lett.} \textbf{\bibinfo{volume}{86}},
  \bibinfo{pages}{1881} (\bibinfo{year}{2001}).

\bibitem[{\citenamefont{Balents et~al.}(2002)\citenamefont{Balents, Fisher, and
  Girvin}}]{Balents02}
\bibinfo{author}{\bibfnamefont{L.}~\bibnamefont{Balents}},
  \bibinfo{author}{\bibfnamefont{M.~P.~A.} \bibnamefont{Fisher}},
  \bibnamefont{and} \bibinfo{author}{\bibfnamefont{S.~M.}
  \bibnamefont{Girvin}}, \bibinfo{journal}{Phys. Rev. B}
  \textbf{\bibinfo{volume}{65}}, \bibinfo{pages}{224412}
  (\bibinfo{year}{2002}).

\bibitem[{\citenamefont{Senthil and
      Motrunich}(2002)}]{Senthil-Motrunich}
  \bibinfo{author}{\bibfnamefont{T.}~\bibnamefont{Senthil}}
  \bibnamefont{and}
  \bibinfo{author}{\bibfnamefont{O.}~\bibnamefont{Motrunich}},
  \bibinfo{journal}{Phys. Rev. B} \textbf{\bibinfo{volume}{66}},
  \bibinfo{pages}{205104} (\bibinfo{year}{2002});
  \bibinfo{author}{\bibfnamefont{O.~I.} \bibnamefont{Motrunich}}
  \bibnamefont{and}
  \bibinfo{author}{\bibfnamefont{T.}~\bibnamefont{Senthil}},
  \bibinfo{journal}{Phys. Rev. Lett.} \textbf{\bibinfo{volume}{89}},
  \bibinfo{pages}{277004} (\bibinfo{year}{2002}).

\bibitem[{\citenamefont{Willett et~al.}(1987)\citenamefont{Willett, Eisenstein,
  Stormer, Tsui, Gossard, and English}}]{Willett-Pan}
\bibinfo{author}{\bibfnamefont{R.}~\bibnamefont{Willett}},
  \bibinfo{author}{\bibfnamefont{J.~P.} \bibnamefont{Eisenstein}},
  \bibinfo{author}{\bibfnamefont{H.~L.} \bibnamefont{Stormer}},
  \bibinfo{author}{\bibfnamefont{D.~C.} \bibnamefont{Tsui}},
  \bibinfo{author}{\bibfnamefont{A.~C.} \bibnamefont{Gossard}},
  \bibnamefont{and} \bibinfo{author}{\bibfnamefont{J.~H.}
  \bibnamefont{English}},
\bibinfo{journal}{Phys. Rev. Lett.}
  \textbf{\bibinfo{volume}{59}}, \bibinfo{pages}{1776} (\bibinfo{year}{1987});
%
\bibinfo{author}{\bibfnamefont{W.}~\bibnamefont{Pan}},
  \bibinfo{author}{\bibfnamefont{J.-S.} \bibnamefont{Xia}},
  \bibinfo{author}{\bibfnamefont{V.}~\bibnamefont{Shvarts}},
  \bibinfo{author}{\bibfnamefont{D.~E.} \bibnamefont{Adams}},
  \bibinfo{author}{\bibfnamefont{H.~L.} \bibnamefont{Stormer}},
  \bibinfo{author}{\bibfnamefont{D.~C.} \bibnamefont{Tsui}},
  \bibinfo{author}{\bibfnamefont{L.~N.} \bibnamefont{Pfeiffer}},
  \bibinfo{author}{\bibfnamefont{K.~W.} \bibnamefont{Baldwin}},
  \bibnamefont{and} \bibinfo{author}{\bibfnamefont{K.~W.} \bibnamefont{West}},
  \bibinfo{journal}{Phys. Rev. Lett.} \textbf{\bibinfo{volume}{83}},
  \bibinfo{pages}{3530} (\bibinfo{year}{1999}).

\bibitem[{\citenamefont{Moore and Read}(1991)}]{Moore91}
\bibinfo{author}{\bibfnamefont{G.}~\bibnamefont{Moore}} \bibnamefont{and}
  \bibinfo{author}{\bibfnamefont{N.}~\bibnamefont{Read}},
  \bibinfo{journal}{Nucl. Phys. B} \textbf{\bibinfo{volume}{360}},
  \bibinfo{pages}{362} (\bibinfo{year}{1991}).

\bibitem[{\citenamefont{Greiter et~al.}(1992)\citenamefont{Greiter, Wen, and
  Wilczek}}]{Greiter92}
\bibinfo{author}{\bibfnamefont{M.}~\bibnamefont{Greiter}},
  \bibinfo{author}{\bibfnamefont{X.~G.} \bibnamefont{Wen}}, \bibnamefont{and}
  \bibinfo{author}{\bibfnamefont{F.}~\bibnamefont{Wilczek}},
  \bibinfo{journal}{Nucl. Phys. B} \textbf{\bibinfo{volume}{374}},
  \bibinfo{pages}{567} (\bibinfo{year}{1992}).

\bibitem[{\citenamefont{Nayak and Wilczek}(1996)}]{Nayak-Read-Fradkin}
\bibinfo{author}{\bibfnamefont{C.}~\bibnamefont{Nayak}} \bibnamefont{and}
  \bibinfo{author}{\bibfnamefont{F.}~\bibnamefont{Wilczek}},
  \bibinfo{journal}{Nucl. Phys. B} \textbf{\bibinfo{volume}{479}},
  \bibinfo{pages}{529} (\bibinfo{year}{1996});
%
\bibinfo{author}{\bibfnamefont{N.}~\bibnamefont{Read}} \bibnamefont{and}
  \bibinfo{author}{\bibfnamefont{E.}~\bibnamefont{Rezayi}},
  \bibinfo{journal}{Phys. Rev. B} \textbf{\bibinfo{volume}{54}},
  \bibinfo{pages}{16864} (\bibinfo{year}{1996});
%
\bibinfo{author}{\bibfnamefont{E.}~\bibnamefont{Fradkin}},
  \bibinfo{author}{\bibfnamefont{C.}~\bibnamefont{Nayak}},
  \bibinfo{author}{\bibfnamefont{A.}~\bibnamefont{Tsvelik}}, \bibnamefont{and}
  \bibinfo{author}{\bibfnamefont{F.}~\bibnamefont{Wilczek}},
  \bibinfo{journal}{Nucl. Phys. B} \textbf{\bibinfo{volume}{516}},
  \bibinfo{pages}{704} (\bibinfo{year}{1998}).

\bibitem[{\citenamefont{Kitaev}(2003)}]{Kitaev97}
\bibinfo{author}{\bibfnamefont{A.~Y.} \bibnamefont{Kitaev}},
  \bibinfo{journal}{Ann. Phys.} \textbf{\bibinfo{volume}{303}},
  \bibinfo{pages}{2} (\bibinfo{year}{2003}), \bibinfo{note}{quant-ph/9707021}.

\bibitem[{\citenamefont{Freedman}(2001)}]{Freedman01}
\bibinfo{author}{\bibfnamefont{M.~H.} \bibnamefont{Freedman}},
  \bibinfo{journal}{Found. Comput. Math.} \textbf{\bibinfo{volume}{1}},
  \bibinfo{pages}{183} (\bibinfo{year}{2001}).

\bibitem[{\citenamefont{Kauffman and Lins}(1994)}]{Kauffman94}
\bibinfo{author}{\bibfnamefont{L.}~\bibnamefont{Kauffman}} \bibnamefont{and}
  \bibinfo{author}{\bibfnamefont{S.}~\bibnamefont{Lins}},
  \emph{\bibinfo{title}{Temperley Lieb Recoupling theory and invariants of
  3-manifolds.}} (\bibinfo{publisher}{Princeton Univ. Press},
  \bibinfo{year}{1994}), vol. \bibinfo{volume}{134} of
  \emph{\bibinfo{series}{Ann. Math. Stud.}}

\bibitem[{\citenamefont{Freedman}(2003)}]{Freedman03}
\bibinfo{author}{\bibfnamefont{M.~H.} \bibnamefont{Freedman}},
  \bibinfo{journal}{Commun. Math. Phys.} \textbf{\bibinfo{volume}{234}},
  \bibinfo{pages}{129} (\bibinfo{year}{2003});
\bibitem[{\citenamefont{Freedman
  et~al.}(2004{\natexlab{a}})\citenamefont{Freedman, Nayak, and
  Shtengel}}]{FNS04a}
\bibinfo{author}{\bibfnamefont{M.}~\bibnamefont{Freedman}},
  \bibinfo{author}{\bibfnamefont{C.}~\bibnamefont{Nayak}}, \bibnamefont{and}
  \bibinfo{author}{\bibfnamefont{K.}~\bibnamefont{Shtengel}}
  (\bibinfo{year}{2004}{\natexlab{a}}), \eprint{cond-mat/0408257}.

\bibitem[{\citenamefont{Greiner et~al.}(2002)\citenamefont{Greiner, Mandel,
  Esslinger, Hansch, and Bloch}}]{Greiner02}
\bibinfo{author}{\bibfnamefont{M.}~\bibnamefont{Greiner}},
  \bibinfo{author}{\bibfnamefont{O.}~\bibnamefont{Mandel}},
  \bibinfo{author}{\bibfnamefont{T.}~\bibnamefont{Esslinger}},
  \bibinfo{author}{\bibfnamefont{T.~W.} \bibnamefont{Hansch}},
  \bibnamefont{and} \bibinfo{author}{\bibfnamefont{I.}~\bibnamefont{Bloch}},
  \bibinfo{journal}{Nature} \textbf{\bibinfo{volume}{415}}, \bibinfo{pages}{39}
  (\bibinfo{year}{2002}).

\bibitem[{\citenamefont{Ioffe et~al.}(2002)\citenamefont{Ioffe, Feigel'man,
  Ioselevich, Ivanov, Troyer, and Blatter}}]{Ioffe02}
\bibinfo{author}{\bibfnamefont{L.~B.} \bibnamefont{Ioffe}},
  \bibinfo{author}{\bibfnamefont{M.}~\bibnamefont{Feigel'man}},
  \bibinfo{author}{\bibfnamefont{V.~A.} \bibnamefont{Ioselevich}},
  \bibinfo{author}{\bibfnamefont{D.}~\bibnamefont{Ivanov}},
  \bibinfo{author}{\bibfnamefont{M.}~\bibnamefont{Troyer}}, \bibnamefont{and}
  \bibinfo{author}{\bibfnamefont{G.}~\bibnamefont{Blatter}},
  \bibinfo{journal}{Nature} \textbf{\bibinfo{volume}{415}}, \bibinfo{pages}{503
  } (\bibinfo{year}{2002}).

\bibitem[{\citenamefont{Freedman
  et~al.}(2004{\natexlab{b}})\citenamefont{Freedman, Nayak, Shtengel, Walker,
  and Wang}}]{Freedman04a}
\bibinfo{author}{\bibfnamefont{M.}~\bibnamefont{Freedman}},
  \bibinfo{author}{\bibfnamefont{C.}~\bibnamefont{Nayak}},
  \bibinfo{author}{\bibfnamefont{K.}~\bibnamefont{Shtengel}},
  \bibinfo{author}{\bibfnamefont{K.}~\bibnamefont{Walker}}, \bibnamefont{and}
  \bibinfo{author}{\bibfnamefont{Z.}~\bibnamefont{Wang}},
  \bibinfo{journal}{Ann. Phys.} \textbf{\bibinfo{volume}{310}},
  \bibinfo{pages}{428} (\bibinfo{year}{2004}{\natexlab{b}}),
  \eprint{cond-mat/0312273}.

\bibitem[{\citenamefont{Freedman
  et~al.}(2002{\natexlab{a}})\citenamefont{Freedman, Larsen, and
  Wang}}]{Freedman02}
\bibinfo{author}{\bibfnamefont{M.~H.} \bibnamefont{Freedman}},
  \bibinfo{author}{\bibfnamefont{M.~J.} \bibnamefont{Larsen}},
  \bibnamefont{and} \bibinfo{author}{\bibfnamefont{Z.}~\bibnamefont{Wang}},
  \bibinfo{journal}{Commun. Math. Phys.} \textbf{\bibinfo{volume}{227}},
  \bibinfo{pages}{605} (\bibinfo{year}{2002}{\natexlab{a}});
%
  \bibnamefont{\emph{ibid,}}
  \textbf{\bibinfo{volume}{228}},
  \bibinfo{pages}{177} (\bibinfo{year}{2002}{\natexlab{b}}).

\bibitem[{\citenamefont{Freedman
  et~al.}(2003{\natexlab{b}})\citenamefont{Freedman, Nayak, and
  Shtengel}}]{FNS03b}
\bibinfo{author}{\bibfnamefont{M.}~\bibnamefont{Freedman}},
  \bibinfo{author}{\bibfnamefont{C.}~\bibnamefont{Nayak}}, \bibnamefont{and}
  \bibinfo{author}{\bibfnamefont{K.}~\bibnamefont{Shtengel}}
  (\bibinfo{year}{2003}{\natexlab{b}}), \bibinfo{note}{cond-mat/0309120}.

\bibitem[{\citenamefont{Roger}(1984)}]{Roger84}
\bibinfo{author}{\bibfnamefont{M.}~\bibnamefont{Roger}},
  \bibinfo{journal}{Phys. Rev. B} \textbf{\bibinfo{volume}{30}},
  \bibinfo{pages}{6432} (\bibinfo{year}{1984}).

\bibitem[{\citenamefont{Herring}(1962)}]{Herring62}
\bibinfo{author}{\bibfnamefont{C.}~\bibnamefont{Herring}},
  \bibinfo{journal}{Rev. Mod. Phys.} \textbf{\bibinfo{volume}{34}},
  \bibinfo{pages}{631} (\bibinfo{year}{1962}).

\bibitem[{\citenamefont{MacDonald et~al.}(1988)\citenamefont{MacDonald, Girvin,
  and Yoshioka}}]{MacDonald88}
\bibinfo{author}{\bibfnamefont{A.~H.} \bibnamefont{MacDonald}},
  \bibinfo{author}{\bibfnamefont{S.~M.} \bibnamefont{Girvin}},
  \bibnamefont{and} \bibinfo{author}{\bibfnamefont{D.}~\bibnamefont{Yoshioka}},
  \bibinfo{journal}{Phys. Rev. B} \textbf{\bibinfo{volume}{37}},
  \bibinfo{pages}{9753} (\bibinfo{year}{1988}).

\bibitem[{\citenamefont{Thouless}(1965)}]{Thouless65}
\bibinfo{author}{\bibfnamefont{D.~J.} \bibnamefont{Thouless}},
  \bibinfo{journal}{Proc. Phys. Soc.} \textbf{\bibinfo{volume}{86}},
  \bibinfo{pages}{893} (\bibinfo{year}{1965}).

\bibitem[{\citenamefont{Chakravarty et~al.}(1999)\citenamefont{Chakravarty,
  Kivelson, Nayak, and Voelker}}]{Chakravarty}
\bibinfo{author}{\bibfnamefont{S.}~\bibnamefont{Chakravarty}},
  \bibinfo{author}{\bibfnamefont{S.}~\bibnamefont{Kivelson}},
  \bibinfo{author}{\bibfnamefont{C.}~\bibnamefont{Nayak}}, \bibnamefont{and}
  \bibinfo{author}{\bibfnamefont{K.}~\bibnamefont{Voelker}},
  \bibinfo{journal}{Phil. Mag. B} \textbf{\bibinfo{volume}{379}},
  \bibinfo{pages}{859} (\bibinfo{year}{1999});
\bibinfo{author}{\bibfnamefont{K.}~\bibnamefont{Voelker}} \bibnamefont{and}
  \bibinfo{author}{\bibfnamefont{S.}~\bibnamefont{Chakravarty}},
  \bibinfo{journal}{Phys. Rev. B} \textbf{\bibinfo{volume}{64}},
  \bibinfo{pages}{235125} (\bibinfo{year}{2001}).

\bibitem[{\citenamefont{Sutherland}(1988)}]{Sutherland88}
\bibinfo{author}{\bibfnamefont{B.}~\bibnamefont{Sutherland}},
  \bibinfo{journal}{Phys. Rev. B} \textbf{\bibinfo{volume}{37}},
  \bibinfo{pages}{3786} (\bibinfo{year}{1988}).

\bibitem[{\citenamefont{Kenyon and R\`{e}mila}(1996)}]{Kenyon96}
  {\bibfnamefont {D.~Jetchev, private communication. For a rigorous
      proof of ergodicity in a similar setting see}}
  \bibinfo{author}{\bibfnamefont{C.}~\bibnamefont{Kenyon}}
  \bibnamefont{and}
  \bibinfo{author}{\bibfnamefont{E.}~\bibnamefont{R\`{e}mila}},
  \bibinfo{journal}{Discrete Math.} \textbf{\bibinfo{volume}{152}},
  \bibinfo{pages}{191} (\bibinfo{year}{1996}).

\end{thebibliography}
\end{document}